\documentclass[aps,pra,twocolumn,showpacs,superscriptaddress]{revtex4}  % for review and submission
\usepackage{graphicx}  % needed for figures
\usepackage{dcolumn}   % needed for some tables
\usepackage{bm}        % for math
\usepackage{amssymb}   % for math
\usepackage{braket}
\usepackage{gensymb}
\usepackage{amssymb,amsmath,amsbsy}
\usepackage[utf8]{inputenc}
\graphicspath{{figures/}}

\begin{document}

\title{Cavity enhanced Raman heterodyne spectroscopy in Er$^{3+}$:Y$_2$SiO$_5$ for microwave to optical signal conversion}

\author{Xavier Fernandez-Gonzalvo}
\author{Sebastian P. Horvath}
\author{Yu-Hui Chen}
\email{stephen.chen@otago.ac.nz}
\author{Jevon J. Longdell}
\email{jevon.longdell@otago.ac.nz}
\affiliation{The Dodd-Walls Centre for Photonic and Quantum Technologies, Department of Physics, University of Otago, 730 Cumberland Street, Dunedin, New Zealand.}

\date{\today}

\begin{abstract}
  The efficiency of the frequency conversion process at the heart of Raman heterodyne spectroscopy was improved by nearly four orders of magnitude by resonant enhancement of both the pump and signal optical fields. Our results using an erbium doped Y$_2$SiO$_5$ crystal at temperatures near 4\,K suggest that such an approach is promising for the quantum conversion of microwave to optical photons.
\end{abstract}


\maketitle

The ability to transfer quantum states encoded in microwave frequency excitations to light would greatly enrich superconducting qubits as a platform for quantum information processing.
Superconducting qubits operate in the microwave frequency regime and can interact strongly with microwave photons \cite{clar2008,scho2008,devo2013,wall04}. However quantum states encoded in microwave photons are swamped by thermal noise at room temperature. This and the relatively high loss of room temperature microwave transmission lines means that microwave photons can't be used for long distance quantum communication. Another issue is that apart from superconducting qubits they interact much more weakly with resonant media such as atoms, ions, or solid state systems than does light. This makes it much more difficult to make quantum memories for microwave photons \cite{clel2004,sill2007,wu2010,timo2011} than to make quantum memories for light \cite{long2005,hoss2009,usma2010,hedg2010,timo2013}. Microwave to optical frequency conversion is also being investigated as a way of protecting classical receivers from overloading by high power surges \cite{ilch02,hsu07}. The idea here is to replace the electronic components of the front end of a microwave receiver with dielectric ones, which are much more robust against high power input signals.

\begin{figure}
  \includegraphics[width=\columnwidth]{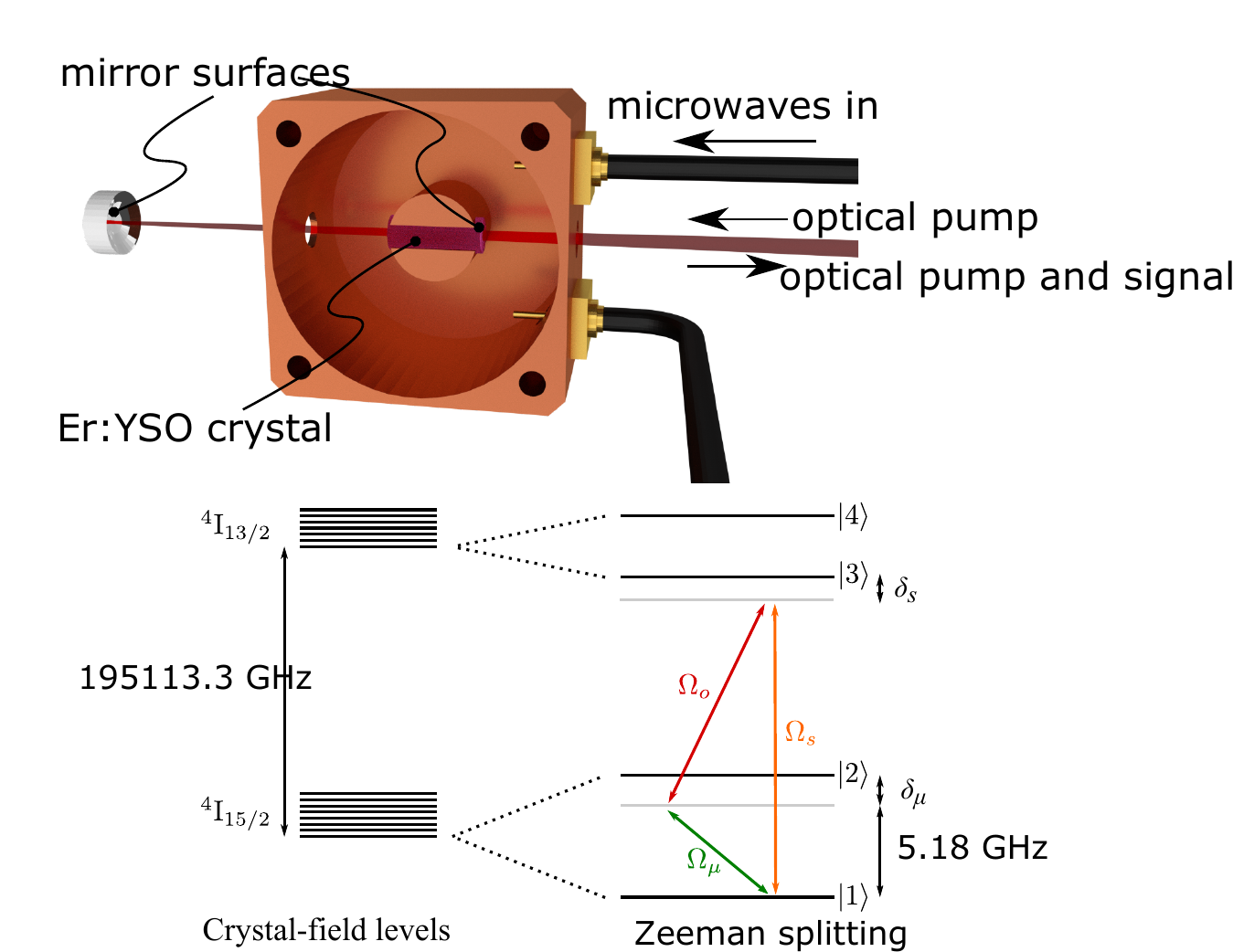}
  \caption{\label{fig:energy_levels}
    \emph{Top:} A render of the upconversion apparatus. A Fabry-Perot resonator is formed between a mirror coated on one end of an Er$^{3+}$:Y$_2$SiO$_5$ crystal and a mirror external to the microwave resonator. The microwave resonator is a loop-gap design. The top of the microwave resonator, which has been removed, contains a circular plunger that forms the other half of the `gaps' in the resontor allowing for broad frequency tunability. Microwaves excite the cavity via an antenna attached to a coax line. A second coax line is used for transmission measurements.
    \emph{Bottom:} Energy level diagram for the even isotopes of Er$^{3+}$:Y$_2$SiO$_5$ showing the transition used to realize upconversion. The crystal field splits the $^4$I$_{15/2}$ and $^4$I$_{13/2}$ levels into 8 and 7 doubly degenerate states (Kramers doublets). Here we use the narrow optical transitions between the lowest doublets of the $^4$I$_{15/2}$ and the $^4$I$_{13/2}$ multiplets, after they have been split by an applied magnetic field.}
\end{figure}

A number of approaches with a wide range of nonlinear frequency mixing mechanisms have been investigated. Optomechanical approaches are the most advanced \cite{andr14,bagc14,boch13} achieving double digit percentage quantum efficiencies with kilohertz bandwiths and with a few added photons of noise. Electroopic approaches \cite{ilch02,hsu07,tsan11} have achieved 0.1\% quantum efficiencies with megahertz bandwiths \cite{rued16} using an optical whispering gallery mode resonator made of lithium niobate. Faraday rotation and ferromagnetic resonance in yttrium iron garnet (YIG) has been explored \cite{hisa16,zhan16,haig16}, as well as using Rydberg atoms \cite{kiff16,han17} and quantum dots \cite{Tsuchimoto17}. This work follows theoretical proposals for microwave-optical conversion using rare earth ions in solids \cite{will14b,obri14}. The approach followed here \cite{will14b} is a cavity enhanced version of Raman heterodyne spectroscopy \cite{mlyn83,wong83}. Raman heterodyne spectroscopy is a method for optically detecting spin resonances. A radio-frequency (RF) signal and an optical pump beam are applied to the sample, and if both the RF and the optical fields are near-resonant with transitions in the material the sum or difference frequency generation can result, as shown in Fig.~\ref{fig:energy_levels}. The generated optical field is in the same mode as the exciting optical field and so easily detected as a beat on the optical pump. Raman heterodyne spectroscopy has been used extensively both to characterise spin energy structures \cite{mits84,mans92,long02prb,long06} and to optically read out spin coherences \cite{eric90,frav04a,frav05}, including the observation of the longest solid-state coherence time \cite{zhon15}. Raman heterodyne spectroscopy was extended into the microwave regime by Schweika-Kresimon et al.\ \cite{schw02} using ruby and has recently been used to show microwave to optical coherent frequency conversion in rare earth doped crystals \cite{fern15}. Here we achieve a quantum efficiency seven orders of magnitude higher than in those initial experiments \cite{fern15}.

In our experiment, a cylindrical sample of erbium doped yttrium orthosilicate (Y$_2$SiO$_5$) of 5\,mm diameter and 12\,mm long sits in the central hole of a shielded loop-gap resonator. The loop-gap resonator has a similar design to that used by Chen et al.\ in \cite{chen16}, where the `gaps' are adjusted using a translation stage enabling broad frequency tuneability. The sample has 10\,ppm of the yttrium ions replaced by erbium, with the natural erbium isotopic ratios. The Y$_2$SiO$_5$ crystal has yttrium ions sitting in two different crystallographic sites and each of these sites has two different orientations, which are related by the crystal's $C_2$ symmetry. In this work we used site~1 \cite{maksimov1971crystal} and made both orientations equivalent by using a magnetic field applied perpendicular to the $C_2$ symmetry axis.

\begin{figure}[!hbt]
  \centering
  \includegraphics[scale=0.5]{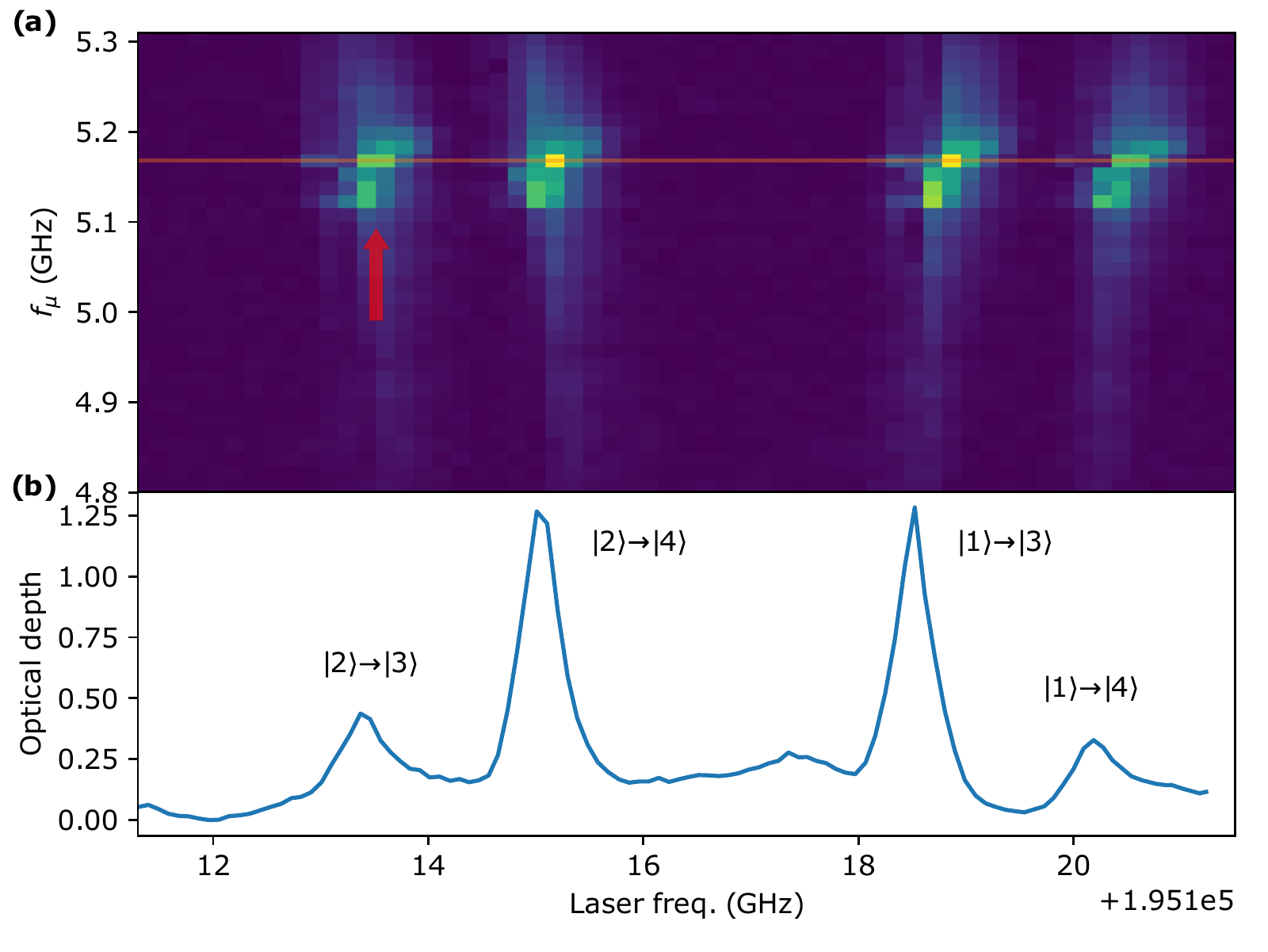}
  \caption{\label{fig:doublepass}%
  (a) Raman-heterodyne signal for Er$^{3+}$:Y$_2$SiO$_5$ in double-pass configuration. The red line indicates the microwave resonance frequency (5.186\,GHz) of the resonator-atom system for the applied magnetic field. The red arrow marks the $\Delta$ system used in this paper. (b) Optical absorption spectrum of the Er$^{3+}$:Y$_2$SiO$_5$ sample for the same external magnetic field. $^{167}$Er (23\%) has nuclear spin of $I=7/2$ and its hyperfine structure is responsible for the broad background.}
\end{figure}

One end of the crystal was cut with a convex shape with a curvature radius of 100.305\,mm and was coated to achieve a reflectivity of 98.8\%. The other end was flat and antireflection coated with a reflectivity lower than 0.8\%. The curved surface and an external highly reflective flat mirror placed outside the microwave resonator form an optical Fabry-Perot resonator. The combined microwave and optical resonators are placed in a 3 axis superconducting magnet inside a cryostat at 4.6\,K. The frequency conversion process uses three energy levels in a $\Delta$ configuration as shown in Fig.~\ref{fig:energy_levels}. By applying an external magnetic field the degenerate Kramers doublets are split into a total of four levels, allowing for four distinct $\Delta$ systems. This is experimentally demonstrated by the four hot spots visible in Fig.~\ref{fig:doublepass}(a). 
%In our experiments we focused on the $\Delta$ system illustrated in Fig.~\ref{fig:energy_levels}, with corresponding optical resonances labeled in Fig.~\ref{fig:doublepass}(b).  \stepcomment{we say this again in the following}

The addition of the optical cavity adds the requirement that the optical pump field as well as the upconverted signal have to be on resonance with an optical cavity mode, which further requires that the microwave frequency is an integer multiple of the optical cavity free spectral range. This requirement is complicated by the fact that the erbium transitions pull the cavity frequencies significantly, which means that the spacing between optical modes is no longer a constant. Moreover changing the magnetic field, which changes the dopants energy levels, can move the cavity resonances as well. To overcome these problems we performed a two dimensional scan of the optical resonator frequency and the microwave driving frequency. While this scan was carried out we locked the pump laser to the optical resonator and kept the microwave resonator in resonance with the input microwave field.

In order to model the conversion process happening in our experiment, the atoms were treated as an ensemble of three level atoms with inhomogeneous broadenings in both the spin and the optical transitions. The cavity modes were treated classically. The atom-cavity system can be described using the following Hamiltonian:
\begin{equation}
\begin{split}
H   & = \sum_{k}(\delta_{\mu,k} \sigma_{22,k} + \delta_{s,k} \sigma_{33,k}) \\ &+ \sum_{k} (\Omega_{o,k} \sigma_{32,k} + g_{s,k} \sigma_{31,k}a + g_{\mu,k} \sigma_{21,k} b + \text{h.c.}).
\end{split}
\label{eq:ham_sys}
\end{equation} 
Here, the sum over $k$ is over all the atoms, $\delta_{\mu,k}$ is the microwave detuning for the $k$th atom, $\delta_{s,k} = \delta_{o,k} + \delta_{\mu,k}$ is the optical detuning of the upconverted optical light, and $\delta_{o,k}$ is the pump laser detuning. $\Omega_{o,k}$ is the Rabi frequency of the pump laser, $\sigma_{ij} \equiv |i \rangle \langle j |$, and $g_{\mu,k}$ and $g_{s,k}$ are the coupling strengths of the $k$th atom to the microwave and optical modes, respectively.

The steady state of the system was found by iteration. First the steady states for the atoms were found for trial cavity amplitudes by solving the atoms' master equations as a function their detunings from the center of the inhomogeneous lines. Second, the atomic coherences were integrated over to update the cavity mode amplitudes. These two steps were repeated until the combined steady state was found.

The parameters for this modelling were all measured directly from the experiment or were a combination of known properties of erbium in Y$_2$SiO$_5$ and modelling of our cavity fields \cite{bott06,thie10}. The least well-constrained parameters were the spin lifetimes and the spin and optical coherence times, which haven't been measured at our particular magnetic field and temperature. We took the spin population lifetime to be 1\,ms and both the optical and spin coherence times to be 1\,$\mu$s. The optical  and microwave inhomogeneous broadenings were 340\,MHz and 50\,MHz respectively. 

In our experiment, an external magnetic field of 146\,mT contained in the $D_1$-$D_2$ plane was applied at an angle of $\sim 29^\circ$ from $D_1$. This angle was chosen to maximise the difference between the quantization axes for the ground and excited state spins. Several preliminary experiments were carried out to characterize the resulting $\Delta$ systems. First, electron paramagnetic resonance was used to find the microwave transition frequency between the two ground states, which was 5.186\,GHz in our case, as indicated by the red line in Fig.~\ref{fig:doublepass}(a). Moving the plunger of the microwave cavity allowed us to tune its frequency and drive the microwave transition on or off resonance. By removing the external mirror of the optical cavity the optical absorption depth was measured, as shown in Fig.~\ref{fig:doublepass}(b). For this measurement, the laser entered the sample from the side normally occupied by the optical cavity external mirror, and was reflected off the mirror coated surface of the Er$^{3+}$:Y$_2$SiO$_5$ crystal. Since the laser makes two passes across the length of the sample, this optical configuration is referred to as the double pass setup. The polarization of the incident light was adjusted so that its electric field was parallel to $D_2$.

Driving a $\Delta$ system with an optical pump and a microwave source, an optical sideband is generated, which can be measured via heterodyne detection \cite{fern15}. Fig.~\ref{fig:doublepass}(a) shows the Raman heterodyne data obtained without the optical cavity for different optical pump and microwave frequencies. Four hot spots corresponding to four $\Delta$ systems can be clearly identified. A closer examination shows that each of the four peaks are actually double peaks. This can be understood by considering the absorption depth of the sample at the upconverted signal frequency. For upconverted photons with a frequency corresponding to the center of the inhomogenous line of $\ket{1} \to \ket{3}$, the absorption depth is at a maximum, and hence they are reabsorbed and contribute less to the heterodyne signal. For photons with slightly off-resonant frequencies this reabsorption is reduced, resulting in two maxima. The maximum quantum efficiency we achieved in the double pass setup was $\eta =2.25\times 10^{-9}$.  In our following experiments using the optical cavity, we focused on the $\Delta$ system with the lowest optical pump frequency, as shown in Fig.~\ref{fig:energy_levels} and marked with a red arrow in Fig.~\ref{fig:doublepass}(a). 

\begin{figure}[!hbt]
  \centering
  \includegraphics[scale=0.65]{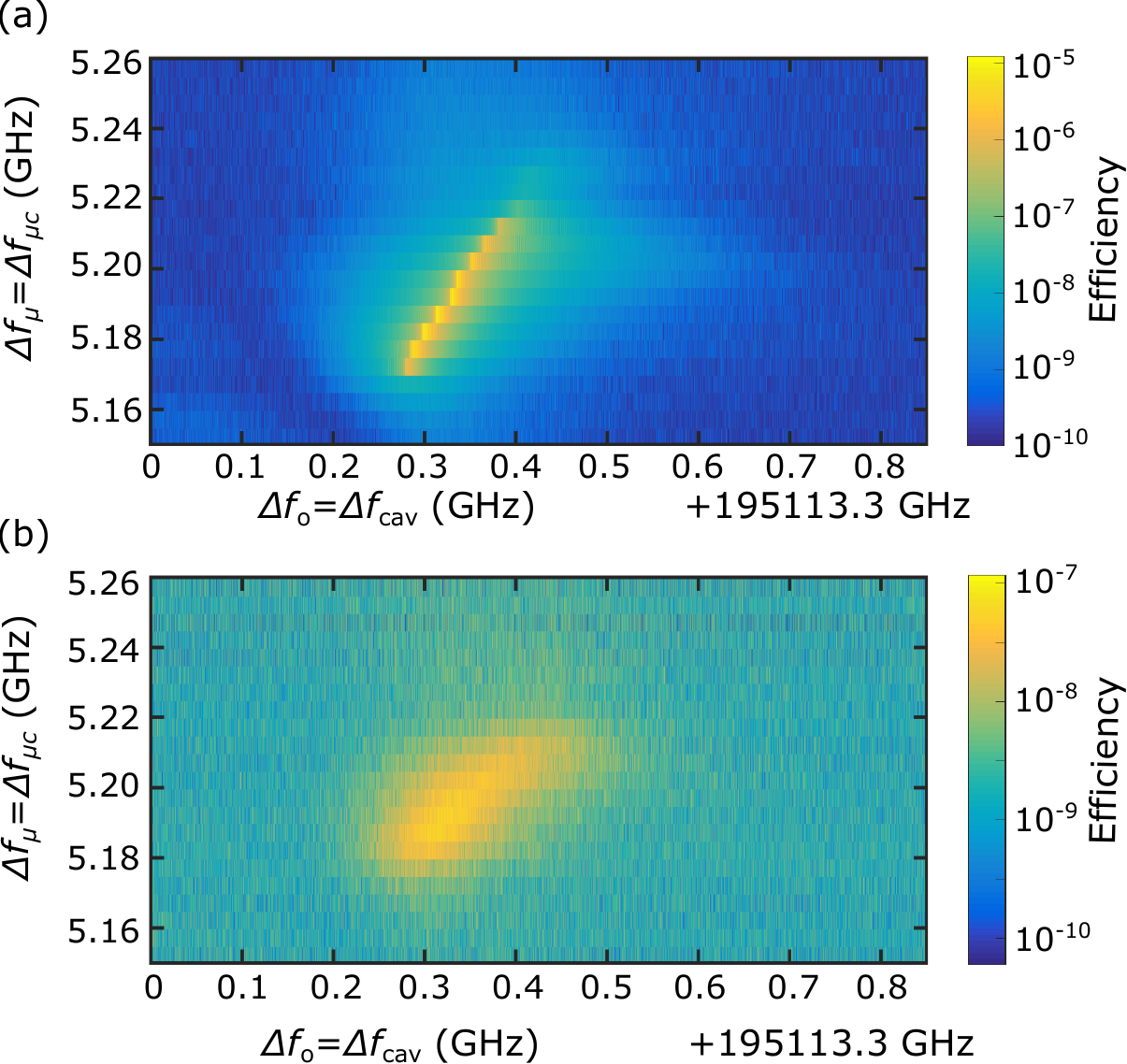}
  \caption{\label{fig:eff_maps}%
  (a) Upconversion efficiency with respect to microwave and laser detuning for a microwave input power of -9.5\,dBm. The maximum efficiency is 1.26$\times10^{-5}$. (b) An identical scan to above, except at a lower microwave input power of -19.5\,dBm. The maximum efficiency is 1.17$\times10^{-7}$. The optical pump power for both measurements was 6.48\,mW. }.
\end{figure}

\begin{figure}[!hbt]
  \centering
  \includegraphics[width=\columnwidth]{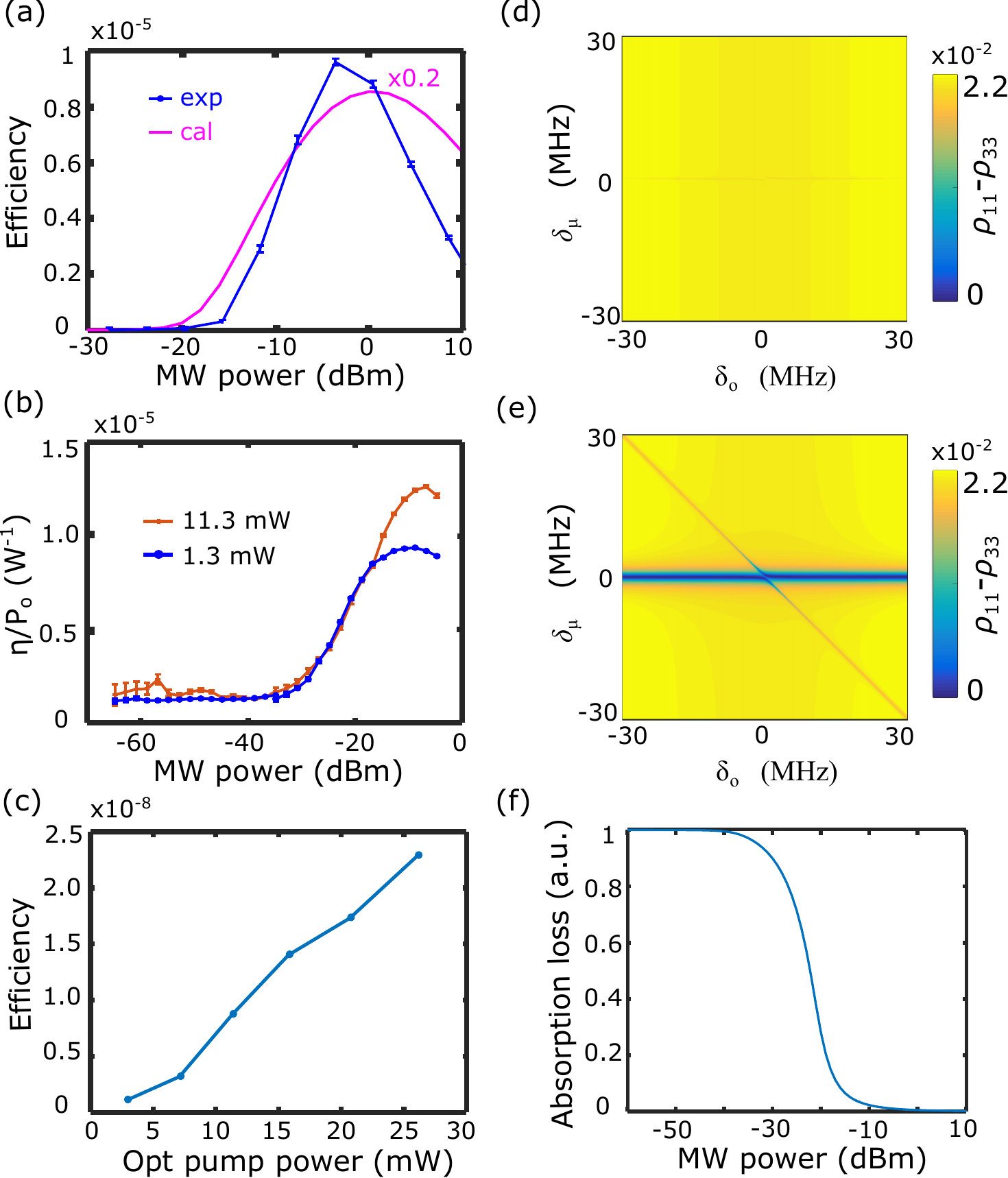}
  \caption{\label{fig:power_dep}%
  (a) Upconversion efficiency as a function of the input microwave power. The experimental data, shown in blue, is for $f_{\text{las}} = 195113.30$\,GHz, $f_{\mu} = 5.18$\,GHz, and a pump laser power of 12\,mW. The efficiencies calculated from our model are shown in magenta. To plot these on the same axes the calculated curve is divided by a factor of five. (b) Efficiency per optical pump power as a function of input microwave power, with the pump laser power as noted. $f_{\text{las}} = 195113.40$\,GHz, $f_{\mu} = 5.18$\,GHz. (c) Upconversion efficiency as a function of the pump laser power. Here the input microwave power was -44.7\,dBm, $f_{\text{las}} = 195113.40$\,GHz and $f_{\mu} = 5.18$\,GHz. (d) Population difference $\rho_{11}-\rho_{33}$ as a function of optical and microwave detunings for an input microwave power of -50\,dBm. (e) Same as (d) but for an input microwave power of -16\,dBm. (f) Absorption loss of the ion ensemble at the upconverted frequency as a function of the microwave power.}
\end{figure}

After introducing the optical cavity, to determine optimal values for the optical pump and microwave detuning parameters, $\delta_o$ and $\delta_{\mu}$, we performed a series of optical and microwave scans at different microwave powers, summarized in Fig.~\ref{fig:eff_maps}. The laser frequency is locked to the optical-cavity resonance frequency, which is scanned across the $\ket{2} \to \ket{3}$ transition shown in Fig.~\ref{fig:doublepass}(b). This system yields a maximum conversion efficiency of $1.26 \times 10^{-5}$ at optimal microwave powers. However when the microwave power is lowered we observe a steep decrease in efficiency. Furthermore, it is evident that the range of laser detunings $\delta_o$ over which a high efficiency is achieved also depends on the microwave power; namely, the range narrows considerably at high powers. To explore this, detailed microwave and optical power dependence measurements were carried out, which are presented in Fig.~\ref{fig:power_dep}. 

Figure~\ref{fig:power_dep}(a) shows both the experimental as well as the theoretical microwave power dependence of the upconversion efficiency. These measurements were performed for a pump laser frequency $f_{\text{las}} =$~195113.30\,GHz and a microwave frequency $f_{\text{mw}}=$~5.18\,GHz. The modelling had no fitted parameters and a good qualitative agreement was achieved. We put down the magnitude discrepancy to the fact that we did not have measurements of the coherence times for the spin, the optical coherence times and lifetimes for these experimental conditions. The model also ignores the spatial variation of both the microwave and the optical fields across the sample as well as the absorption of $^{167}$Er. (We note that the maximum efficiency in Fig.~\ref{fig:power_dep}(a) is $0.95 \times 10 ^{-5}$, which is slightly lower than that shown in Fig.~\ref{fig:eff_maps}(a). This discrepancy is because the narrowing of the high-efficiency microwave region combined with a small drift of the optical cavity during these measurements resulting in a $\delta_o$ slightly off-set from maximum efficiency.)

By plotting the conversion efficiency per unit of pump power ($\eta/P_o$) as a function of microwave power for two optical pump powers, Fig~\ref{fig:power_dep}(b) shows an approximately linear scaling of the conversion efficiency with respect to the optical pump power. This linear relationship is more directly established in Fig.~\ref{fig:power_dep}(c), for which the efficiency was measured as a function of the optical pump power using a fixed microwave power of $-44.7$ dBm.

To understand the strong dependence on microwave power, we note that for a microwave frequency of $5.18$ GHz, the upconverted photons have a frequency which falls within the inhomogeneous linewidth of the $\ket{1} \to \ket{3}$ transition. Consequently, a large fraction of the upconverted photons are reabsorbed prior to exiting the optical cavity. To understand this we need to consider how the population distributions react as the microwave power is changed.

For given optical and microwave input frequencies the ions satisfying the condition $\delta_o + \delta_{\mu}=0$ will be on resonance with the generated sideband, and thus reabsorb the upconverted signal. This reabsorption will be greatly enhanced by the presence of the optical cavity. The magnitude of the reabsorption process will be determined by the population difference $\rho_{11}-\rho_{33}$ between energy levels $\ket{1}$ and $\ket{3}$. Figs.~\ref{fig:power_dep}(d) and~\ref{fig:power_dep}(e) show a calculation of $\rho_{11}-\rho_{33}$ for two different microwave powers (-50\,dBm and -16\,dBm respectively). At low microwave powers the population difference is mostly flat, with variations on the order of $\sim2.2\times 10^{-2}$. In contrast, at higher microwave powers the population difference for ions satisfying $\delta_o + \delta_{\mu}=0$ is lower than for those that don't, resulting in a lower reabsorption of the upconverted light and therefore a higher conversion efficiency. Then, the dependence of the conversion efficiency with the microwave power can be understood as a redistribution of the electronic population in the energy levels, minimizing absorption at the generated sideband frequencies for higher microwave powers. Fig.~\ref{fig:power_dep}(f) shows the prediction for this reduction in absorption, occurring at approximately -20\,dBm, which agrees with the efficiencies plotted in Fig.~\ref{fig:power_dep}(a). The fact that the ions for which $\delta_o + \delta_{\mu} = 0$ correspond to a narrow subset of the inhomogeneous lines explains the apparent narrowing of the conversion efficiency at high microwave powers shown in Fig.~\ref{fig:eff_maps}. The absorption of $^{167}$Er, visible in Fig.~\ref{fig:doublepass}, is ignored by our model but will also affect our signal in a similar way.  As is the case with the even isotopes, this absorption can be saturated at the high signal powers caused by high microwave inputs. 

In order to mitigate the effect of absorption at the upconverted frequency there are a number of strategies that one can consider.
Performing these experiments using isotopically purified samples containing no $^{167}$Er, this reabsorption could be substantially reduced.
 Alternatively, cooling the sample to mK temperatures would result in almost all of the ground-state hyperfine levels being depopulated, leaving only a small number of very narrow hyperfine absorption features. Thus, by appropriately choosing $\delta_o$ and $\delta_{\mu}$, the upconverted frequency could be chosen to avoid reabsorption.

Lowering the temperature sufficiently to freeze out microwave frequency excitations also alters several other aspects of the upconversion scheme. In particular, with all ions in the ground state $\ket{1}$, both the microwave resonator and the optical cavity will be operating in the strong coupling regime. This automatically forces both the input microwave frequency and the optical output frequency off atomic resonances. Operating at mK temperatures boosts the conversion efficiency in two further ways. Firstly, with all ions in the ground state $\ket{1}$, a greater number of ions will be active in the Raman process. Secondly, the effect of dephasing mechanisms that dominate at 4.6\,K is substantially reduced at lower temperatures, increasing the coherence times for both the optical and spin transitions \cite{bott06}.

Aside from this, the conversion efficiency could also be improved by modifying both the microwave resonator and the optical cavity to achieve optimal impedance matching \cite{will14b}. The microwave resonator used in these experiments had coupling losses of $\kappa_{1,\mu} = 75$ kHz, $\kappa_{2,\mu} = 55$ kHz, with an intrinsic loss $\kappa_{i,\mu} = 717$ kHz, yielding a resultant reflection of 76\%. By modifying the antennas this could be reduced to achieve almost perfect impedance matching, corresponding to a boost in efficiency by a factor of $\sim 4.2$. Similarly, the optical cavity in our scheme was over-coupled, with a coupling loss of $\kappa_{1,o} = 8.0$\,MHz and an intrinsic loss of $\kappa_{i,o}=1.7$\,MHz. 

To obtain an estimate of the efficiency for an optimal implementation of the presented scheme, we ran our theoretical model, with our current cavity parameters, assuming ideal impedance matching and a 50 mK working temperature, which yielded a number efficiency of 14\% in the limit of low microwave powers. To further improve the current scheme, reducing the intrinsic losses (as opposed to coupling losses) of the cavities and optimizing the erbium concentration in the sample to obtain stronger nonlinearities can push the efficiency closer to unit. %While several alternate protocols exist for microwave to optical photon converters that achieve comparable efficiencies, operating these system at mK temperatures remains an unsolved problem to date \cite{andr14,bagc14,boch13,rued16}. These results suggest there is great potential for an optical cavity enhanced rare-earth upconversion scheme for hybrid quantum computing systems.

In summary, we have demonstrated microwave to optical-telecommunications-band signal conversion with a number efficiency of $\eta = 1.26\times 10^{-5}$ by means of an Er$^{3+}$:Y$_2$SiO$_5$ crystal embedded in both a microwave cavity and an optical cavity at 4.6\,K. The introduction of the optical cavity in the system has provided an enhancement in the conversion efficiency of $6\times10^3$ compared to its counterpart without an optical cavity. The dependencies on input microwave power and optical pump power give more information about the conversion process. By analysing the experimental data with a theoretical model we propose that matching the impedances of both cavities and lowering the temperature to mK will bring our system to a near unit conversion efficiency.
These results suggest there is great potential for cavity enhanced upconversion in rare earth ions to act as a bridge between microwaves and light for hybrid quantum computing systems.

We would like to acknowledge the Marsden Fund (Contract No. UOO1520) of the Royal Society of New Zealand. We thank Quinn Thorsness for his help in producing Fig.~1.

%\bibliography{jevonrefs,jevonrefs2,sebastianrefs,xavirefs,stephenrefs}

\end{document}